\documentclass[aps,pre,twocolumn,superscriptaddress,superscriptreference]{revtex4-1}

\usepackage{bm,amssymb,graphicx,algorithm,mathtools,scalerel,xcolor}
 
\usepackage{comment}

\makeatletter
\newcount\c@MaxMatrixCols \c@MaxMatrixCols=10

\makeatother

\newcommand{\ud}{\text{d}}
\newcommand{\ui}{\text{i}}

\newcommand{\FS}[1]{{\color{black} FS: #1}}
\newcommand{\fs}[1]{{\color{black} #1}}
\newcommand{\gj}[1]{{\color{black} #1}}
\newcommand{\GJ}[1]{{\color{black} #1}}

\newcommand{\fsremove}[1]{}

\newcommand{\FDT}{\fs{FDT }}
\newcommand{\FDTs}{\fs{FDTs }}
\newcommand{\FDTNS}{\fs{FDT}}
\newcommand{\FDTsNS}{\fs{FDTs}}

\begin{document}

\title[Fluctuation-Dissipation Relations Far from Equilibrium]{Fluctuation-Dissipation Relations Far from Equilibrium: A Case Study}

\author{Gerhard Jung}
\email{gerhard.jung@uibk.ac.at}
\affiliation{Institut f\"ur Theoretische Physik, Universit\"at Innsbruck, 
Technikerstra{\ss}e 21A, A-6020 Innsbruck, Austria}

\author{Friederike Schmid}
\email{friederike.schmid@uni-mainz.de}
\affiliation{Institut f\"ur Physik, Johannes Gutenberg-Universit\"at Mainz, 55099 Mainz, Germany}

\begin{abstract}
Fluctuation-dissipation relations or ``theorems'' (\FDTsNS) are fundamental for
statistical physics and can be rigorously derived for equilibrium
systems. Their applicability to non-equilibrium systems is, however,
debated.  Here, we simulate an active microrheology experiment, in
which a spherical colloid is pulled with a constant external force
through a fluid, creating near-equilibrium and far-from-equilibrium
systems. We characterize the structural and dynamical properties of
these systems, and reconstruct an effective generalized Langevin
equation (GLE) for the colloid dynamics. Specifically, we test the
validity of two \FDTsNS: The first \FDT relates the non-equilibrium
response of a system to equilibrium correlation functions, and the
second \FDT relates the memory friction kernel in the GLE to the
stochastic force. We find that the validity of the first \FDT depends
strongly on the strength of the external driving: it is fulfilled
close to equilibrium and breaks down far from it. In contrast, we
observe that the second \FDT is always fulfilled.  We provide a
mathematical argument why this generally holds for memory kernels
reconstructed from a deterministic Volterra equation for correlation
functions, even for non-stationary non-equilibrium   systems.
 
Motivated by the Mori-Zwanzig formalism, we therefore suggest to
impose an orthogonality constraint on the stochastic force, which is
in fact equivalent to the validity of this Volterra equation. Such
GLEs   automatically satisfy the second \FDT and are unique, which is
desirable when using GLEs for coarse-grained modeling.

\end{abstract}

\maketitle

\section{Introduction}
\label{Sec:Intro}

Fluctuation-dissipation theorems (\FDTsNS) combine
the distinct worlds of ``thermal fluctuations'' and ``dissipative
response'' and have become a cornerstone of statistical physics
\cite{Onsager1931A,Onsager1931B,kubo1966fluctuation,MARCONI2008,Prost2009,Speck2010_FDTNESS,stratonovich2012nonlinear}
with many applications in condensed matter physics
\cite{Gittes1997,Sciortino2001,Berthier2002, JF2007,Krueger2009} (just
to name a few). In the literature several distinct forms of \FDTs
appear. The most common one is derived from linear response theory and
relates the \emph{non-equilibrium} response function of an observable
to the relaxation of \emph{equilibrium} fluctuations. This relation
corresponds to Onsager's hypothesis, stating that a system cannot
differentiate between forced and spontaneous fluctuations
\cite{Onsager1931B}.  In the following this relation will be referred
to as first fluctuation-dissipation relation 1\FDTNS. Another \FDT
\fs{appears in}
\gj{generalized Langevin equations} and connects the systematic,
friction interactions in the system, described by the memory
kernel, with the coloured thermal noise. We refer to this relation as
second fluctuation-dissipation relation 2\FDTNS. 

\GJ{ For equilibrium systems, the \FDTs can be rigorously derived within linear response theory \cite{kubo1966fluctuation,forster1975hydrodynamic}, their validity in non-equilibrium situations
has, however, been extensively and controversially discussed in the literature
\cite{Sciortino2001,Berthier2002,Crisanti_2003,JF2007,Tanaka2009,Grigera2009,Speck2009_FDTNESS,Lippiello2014,Falasco2014,
Maes2014, Steffenoni2016, Zaccone2018_FDTNESS,Vulpiani2019,
Netz2020_FDTNESS,Santamaria2020}. Outside the linear response regime, these theorems should therefore be rather seen as unproven ``relations'' \footnote{\GJ{To avoid confusion we will use the abbreviation FDT to refer to the fluctuation-dissipation theorem applied to non-equilibrium systems, 
although, stictly speaking, theorems cannot be violated}}.} One reason for the controversies
might be that an apparent violation of the \FDT could be caused by an
incorrect generalization of the equilibrium \FDT to non-equilibrium
systems. For example, in the case of active microrheology, it has been
shown that \gj{close to equilibrium} a \FDT can be recovered when
considering an additive correction accounting for the local mean
velocity of the particle
\cite{Agarwal1972_FDTNESS,Speck2009_FDTNESS,Speck2010_FDTNESS}. For
our system this \gj{implies} that the 1\FDT is valid in the Galilean
reference frame that moves with the average velocity of the colloid
(which will be called ``colloid frame'' in the following). This can be
intuitively understood from Onsager's hypothesis according to which
the relaxation of forced fluctuations \emph{in the non-equilibrium
steady-state} should be related to spontaneous fluctuations
\emph{around} this non-equilibrium state. Other situations that can
lead to apparent violation of the 1\FDT have been discussed in
\cite{Sciortino2001,Villamaina2009_FDTTHERMO,Villamaina2012_FDTTHERMO,Vulpiani2019}.
An intuitive Gedanken-experiment is a system in which two thermostats
with different temperature act on different degrees of freedom of a
particle (i.e. in different dimensions) and a cross-correlation exists
between these degrees of freedom.  Such systems appear to violate the
1\FDTNS, however, differences between response and fluctuations can
be directly related to the temperature difference and the strength of
the cross correlations
\cite{Villamaina2009_FDTTHERMO,Villamaina2012_FDTTHERMO}. 
 
Discussions of the 2\FDT in non-equilibrium systems have so far been
scarce in the literature.  From a theoretical perspective, the
situation is clear for dynamical systems with unitary time evolution.
This includes classical and quantum mechanical Hamiltonian systems,
but also quasi-Hamiltonian systems \fs{such as Molecular Dynamics models
that include Nose-Hoover thermostats}.
Applying the Mori-Zwanzig formalism, one can then \emph{exactly} rewrite the
microscopic equations of motion in terms of a GLE for coarse-grained
variables, and derive an 2\FDT for stationary \cite{Zwanzig2001} or
non-stationary \cite{Meyer2017} systems without any assumptions --
apart from the requirement that the space of dynamical variables forms
a Hilbert space (and thus an inner product is defined). In
non-Hamiltonian systems, however, the validity of the 2\FDT has been
questioned, and in fact several recent papers have suggested violation
of the 2\FDT \cite{Falasco2014, Maes2014, Steffenoni2016,
Srivastava2018, Zaccone2018_FDTNESS, Netz2020_FDTNESS}. It is
therefore desirable to understand potential origins for the violation
of \FDTs in non-equilibrium systems.

\fs{\em Note: After publication of this paper, we became aware of two
recent preprints by Zhu et al. \cite{Zhu2021a,Zhu2021b} who applied the
Mori formalism to stochastic systems at equilibrium and non-equilibrium steady states. They
derived a generalized 2FDT using properties of the Kolmogorov operator,
which reduces to the classical 2FDT in many cases, 
including  explicitly any systems where the relevant variables are degenerate (i.e. no direct application of white noise such as the system studied in this work) . }
 
 In \fs{the present} paper, we investigate the validity of the
 fluctuation-dissipation relations in non-equilibrium steady-states
 using the example of active microrheology \cite{Puertas_2014}. For
 this purpose we study the linear and non-linear response of a colloid
 immersed in a fluid described by dissipative particle dynamics (DPD)
 \cite{Hoogerbrugge_1992,Espanol_1995} to an externally applied
 driving force. To evaluate the \FDTs we analyze the properties of the
 tracer particle in the colloid frame in detail. We reconstruct the
 memory kernel, which allows us to \fsremove{unambiguously} determine
 the coloured thermal noise\fsremove{ \emph{without} application of
 the 2\FDT}. In this way we can not only validate the 2\FDT but also
 extract the noise distribution which shows an unexpected, asymmetric
 non-Gaussian behaviour for systems far away from equilibrium (i.e.
 pulling forces outside the linear response regime). Furthermore, we
 observe an apparent violation of the 1\FDT \gj{far away from
 equilibrium}, which we interpret in terms of the aforementioned
 two-thermostat model. 
 
 Our manuscript is organized as follows. In Chapter \ref{Sec:Methods}
 we introduce in detail the two fluctuation-dissipation relations that
 will be studied in this work \gj{and present some novel results on
 the 2\FDT in non-equilibrium \fs{and possibly even} non-stationary
 systems}. \gj{ We then describe the simulation model and
 \fs{analysis} techniques, including the reconstruction of the memory
 kernel and determination of the noise in Chapter
 \ref{sec:simulation_model}.} Afterwards, in Chapter
 \ref{Sec:Response}, we analyze the response of the colloid in the
 reference frame.  The main results of this paper about the properties
 of fluctuations and dissipation in non-equilibrium steady states, as
 well as the validity and breaking of fluctuation-dissipation
 relations are presented in Chapter \ref{Sec:FDR}. In Chapter
 \ref{Sec:Discussion} we then discuss the implications of these
 results for future investigations of non-equilibrium systems. We
 summarize and conclude in Chapter \ref{Sec:Conclusion}.

\section{Fluctuation-Dissipation Relations}
\label{Sec:Methods}

\gj{In this chapter we first review the basic principles of linear
response theory to derive the (first) fluctuation-dissipation relation
(1\FDTNS). Since this formalism can be found in standard textbooks (see,
e.g. Ref.~\cite{forster1975hydrodynamic}) we keep our discussion to a
very minimum, only introducing the fundamental equations that will be
important for the results of this work. In the second part we then
discuss the generalized Langevin equation and how it can be connected
to the second fluctuation-dissipation relation (2\FDTNS), even in
non-stationary non-equilibrium situations.}

\subsection{Linear response theory and the 1\FDT }
\label{Sec:Linear_response}

The fundamental idea of linear response theory is to determine the
time-dependent response function, $ \chi(t) $, which defines the
response of an observable in the system to an external perturbation, $
\alpha(t), $ of the Hamiltonian, $ H = H_0 - \alpha(t) X $. Here, $
H_0 $ is the equilibrium Hamiltonian. Under the assumption that $
\alpha(t) $ is a small parameter and the system is in equilibrium for
$ t <0 $ one can immediately derive the response of an observable $ Y
$, determined by,
\begin{equation}\label{eq:lr}
\delta \langle Y(t) \rangle = \int_{-\infty}^{t} \ud s \chi(t-s) \alpha(s).
\end{equation}
The response function is determined by the 1\FDTNS, which in classical systems can be derived as,
\begin{equation}\label{eq:1FDR}
\chi(t) = - \beta \frac{\ud}{\ud t}C_{XY}(t) \: \Theta(t),
\end{equation}
with the Heaviside $\Theta$ function,
where $ \beta $ is the inverse temperature $ \beta^{-1} = k_B T $ and $ 
C_{XY}(t) $ is the \emph{equilibrium} correlation function,
\begin{equation}
C_{XY}(t) = \left\langle X(0)^* Y(t) \right\rangle_\text{eq} = \left\langle X(0) | Y(t) \right\rangle,
\end{equation}
given by the inner product in the vector space of observables,
\begin{equation}\label{eq:inner_product}
\left\langle X | Y \right\rangle= \int \ud \Gamma \rho_\textrm{eq}(\Gamma) X(\Gamma)^* Y(\Gamma),
\end{equation}
with probability density $ \rho_\textrm{eq}(\Gamma) $ defined on the phase space points $ \Gamma = (x,p) $. 

In Section~\ref{sec:1FDR_vio} we apply a perturbation $ \alpha(t) = M
V_0 \delta(t) $, i.e. an instantaneous force, acting on the position
of the colloid, $ X $, and we investigate the response of the
velocity, $ Y(t) = V(t). $ Here, $ M $ is the colloid mass and $ V_0 $
the instantaneous velocity, $ V_0 = \delta \langle V(0) \rangle $. The
1\FDT can thus be transformed to,
\begin{eqnarray}
\frac{\delta \langle V(t) \rangle}{V_0} &=&  M \chi_V(t) = - \beta M  
\frac{\ud}{\ud t}\left\langle X(0) V(t) \right\rangle_\text{eq} \nonumber 
\\
&=& \beta M  \left\langle \dot{X}(-t) V(0) \right\rangle_\text{eq} \nonumber\\
&=& \beta M  \left\langle V(0) V(t) \right\rangle_\text{eq} \label{1FDR_general} \\
&=& \frac{C^\text{eq}_V(t)}{C^\text{eq}_V(0)},\label{1FDR_normalized}
\end{eqnarray}
with the velocity auto-correlation function (VACF), $ C^\text{eq}_V(t)
= \left\langle V(0) V(t) \right\rangle_\text{eq} $. As has been
discussed in the literature, \gj{under mild assumptions} one expects
this relationship to hold \gj{even in} non-equilibrium steady-states
if the dynamics are investigated in the colloid frame
\cite{Agarwal1972_FDTNESS,Speck2009_FDTNESS,Speck2010_FDTNESS}. \gj{
These assumptions include that the solvent has the same properties as
in equilibrium (i.e. Boltzmann-distributed velocities according to
temperature $  T $) which implies that the system is close to
equilibrium}. In this work, we will apply the perturbation $\alpha(t)$
in stationary but non-equilibrium systems induced by a permanent
external pulling force $F_\text{ext}$ acting on the colloid. The
velocity $V_0$ will be chosen parallel to $F_\text{ext}$. The
assumption \gj{of being close to equilibrium} is thus no longer valid
in situations where the external driving on the colloid is strong
enough to heat up the surrounding fluid. \gj{In this case the
equilibrium averages have to be replaced by non-equilibrium averages,
$ C^\text{ss}_V(t) = \left\langle V(0) V(t) \right\rangle_\text{ss}, $
in the stationary state.} Throughout this work we will identify the
value  \gj{$ T_\textrm{neq} = \frac{C^\text{ss}_V(0) M}{k_B} $ as the
non-equilibrium ``temperature'' of the fluid}. At equilibrium, one
clearly has $ T_\textrm{neq} = T $ (and thus $ \beta_\textrm{neq} =
\beta $). In situations \gj{close to equilibrium} where $
\beta_\textrm{neq} \approx \beta $, Eq.~(\ref{1FDR_general}) can still
assumed to be valid \cite{Speck2010_FDTNESS}. For larger external
driving, Eq.~(\ref{1FDR_normalized}) is correct for $ t=0 $ and it
remains to be investigated whether it also holds for larger times $
t>0 $.

\subsection{\gj{Generalized Langevin Equation} and the 2\FDT }
\label{ref:MZ}

In active microrheology, one usually solely investigates the motion of
the immersed colloid, given by its position and velocities $
\{X(t),Y(t)\} $. The other degrees of freedom in the system, i.e. the
positions and velocities of the solvent particles are thus not
considered, and only affect the motion of the colloid indirectly via
effective equations of motion. If the microscopic dynamics are
Hamiltonian, the Mori-Zwanzig projection operator (MZ) formalism is a
powerful tool to derive an \emph{exact} relation for these effective
dynamics \cite{Zwanzig1961, Mori1965,Zwanzig2001}. \gj{The final
result is given by the generalized Langevin equation (GLE) for 
\fs{a given} set of selected variables $ \{A_i\} $,
\begin{equation}
\frac{\ud}{\ud t} A_i(t) = \ui \Omega_{ij} A_j(t) - \int_{0}^{t} \ud s K_{ij}(t-s) A_j(s) + \partial F_i(t),
\end{equation}
including the frequency matrix $ \Omega_{ij} $ \fs{that describes
direct interactions between the variables $A_i$}, the memory kernel $
K_{ij}(t) $ and the fluctuating force $ \partial F_i(t) $, for which
the MZ formalism provides explicit expressions \cite{Zwanzig2001}, but
which are very difficult to evaluate analytically in a general
framework. From the MZ formalism it is, however, possible to derive a
Volterra equation of first kind, 
\begin{equation}\label{eq:Volterra}
\frac{\ud}{\ud t} {C}_{ij}^\text{eq}(t) = \textrm{i} \Omega_{ik} C^\text{eq}_{kj}(t)  - \int_{0}^t \ud s K_{ik}(t-s) C_{kj}^\text{eq}(s),
\end{equation}
where the correlation function $C^\text{eq}_{ij}(t) = \langle A_i(t)
A_j(0) \rangle_\text{eq} $ is accessible in computer simulations
\cite{Shin2010} and experiments \cite{franosch2011resonances}. This
Volterra equation thus allows to systematically determine the
deterministic parts of the generalized Langevin equation, and it
directly follows from the orthogonality condition for the fluctuating
force,
\begin{equation}\label{key2}
\langle   A_i(0) \partial F_j(t)\rangle_\text{eq} = 0.
\end{equation} 
  }

The expression for the memory kernel in the MZ formalism can also be
transformed into the 2\FDTNS,
\begin{equation}\label{eq:2FDR}
 \left\langle   \partial F_i(0)  \partial F_j(t)  \right\rangle_{\text{eq}} = K_{ik} (t) \left\langle A_k(0) A_j(0)\right\rangle_{\text{eq}},
\end{equation}
which, similar to the 1\FDTNS, therefore directly connects the
friction interactions in the system with \fs{the correlations of}
fluctuations. \fs{We should however note that in general, the MZ
formalism does not predict the fluctuating forces to be Gaussian
distributed, and indeed, strong deviations from Gaussianity have
been observed even in simple equilibrium systems \cite{Shin2010,carof2014two}}.
\gj{An important application of the 2\FDT
is for example the Nyquist relation, which relates the resistance of a
resistor to its thermal electric noise
\cite{balakrishnan1979fluctuation,Nyquist1928}. The 2\FDT also plays
an important role in non-Markovian modeling
\cite{Shin2010,franosch2011resonances,Li2015}.}

\gj{As has been discussed in the introduction, the validity of this
2\FDT in dissipative systems far from equilibrium has been questioned.
However, \fs{we will now show} that it is much more generally valid. 
\fs{For simplicity, we omit direct interactions between
selected variables in the following.} 
\fs{We consider a set of selected
variables $A_i(t)$ whose dynamical evolution is characterized by a
correlation matrix $C_{ij}(t,t_0) = \langle A_i(t) A_j(t_0)
\rangle_\text{neq} $. Here, $\langle ...  \rangle_\text{neq}$ denotes
the non-equilibrium average over an ensemble of trajectories starting
from an initial probability density $\rho(\Gamma)$ at an arbitrary
''initial'' time $T<t,t_0$, which can also be chosen $T \to - \infty$.
We do not impose invariance with respect to time translation.
However, we assume that the correlation functions can be connected to
memory kernels $K_{ij}(t,t_0)$ by means of a deterministic Volterra
equation
\begin{equation}
\label{eq_volterra_general} 
\frac{\ud}{\ud t}
C_{ij}(t,t_0) = - \int_{t_0}^t \ud s \: K_{ik}(t,s) \: C_{kj}(s,t_0).
\end{equation} 
In time-translation invariant systems with $C_{ij}(t,t_0)=C_{ij}(t-t_0)$, 
this is certainly true, as one can solve Eq.\ (\ref{eq_volterra_general}) 

for $K_{ij}(t-s)$ in a straightforward manner using Fourier methods
(with some adjustments in case $\dot{C}_{ij}(0) \neq 0$, 
see Appendix \ref{app:2fdr}).}  
\fs{Eq.~(\ref{eq_volterra_general}) has been derived for Hamiltonian systems
with a time-dependent projection operator formalism by Meyer et
al \cite{Meyer2017}. Here, we take a more general point of view, and
see the equation simply as a way to reparametrize the correlation
functions $C_{ij}(t,t_0)$.  }

\fs{Based on Eq.\ (\ref{eq_volterra_general}), we show in the Appendix
\ref{app:2fdr} that the correlation structure defined by
$C_{ij}(t,t_0)$ can be reproduced by a coarse-grained non-stationary
GLE model of the form \cite{stella2014generalized,Meyer2017},
\begin{equation}
\label{eq:gle_general}
 \frac{\ud }{\ud t} A_i(t) =  
    - \int_{T}^{{t}} \ud s K_{ij}(t,s) A_j(s) + \partial F_i(t),
\end{equation}
where the Volterra equation {\em automatically} implies the 2\FDT
	\begin{equation}
	\langle \partial F_i(t) \: \partial F_j(t_0) \rangle_\text{neq}
	=  \: K_{ik}(t,t_0) \: C_{kj}(t_0,t_0).
	\end{equation}}
} 
\gj{This is a central result of this paper since it states that there
is no fundamental violation of the 2\FDT in non-equilibrium systems.
This statement is not restricted to Hamiltonian systems, and the
derivation does not rely on the Mori-Zwanzig formalism. In the
following we will therefore refer to it as the second
fluctuation-dissipation \emph{theorem} (2\FDTNS) also in non-equilibrium
settings.} 

\gj{It should be noted, that it is possible to establish a relation
to the Mori-Zwanzig framework by noting (see Appendix \ref{app:2fdr})
that Eq.\ (\ref{eq_volterra_general}) is also equivalent to the
requirement $\langle \partial F_i(t) A_j(T) \rangle = 0$ at time
$t_0=T$. Hence the Volterra equation also implies that the fluctuating
force is perpendicular to the selected variable $A_i$ at some
(arbitrary) reference time $T$.}
\GJ{We emphasize that we do not assume the fluctuating force to be
Gaussian distributed.}
	
    In the following, we will first investigate the implications of
    this result on the concrete example of active microrheology. We
    therefore describe the dynamics of the colloid in the colloid
    frame using the selected variables $ \{ A_1=X, A_2 = V \} $,
    resulting in the GLE,
	\begin{align}
	\frac{\ud}{\ud t} X(t) =& V(t),\\
	M\frac{\ud}{\ud t} V(t) =&  - \int_{0}^{t} \ud s \: K(t-s) V(s) + \partial F(t),
	\end{align}
	together with the 2\FDTNS,
	\begin{equation}\label{eq:fdr2_first}
	\gj{\left\langle   \partial F(0) \partial F(t)  \right\rangle_\text{ss} = k_B T_\text{neq} K (t).}
	\end{equation}
\gj{	The only difference to the equilibrium case is thus the usage of the 
non-equilibrium temperature $ k_B T_\text{neq} = {C^\text{ss}_V(0) M} $, as defined above.}
	
In recent years several different numerical algorithms have been proposed 
to calculate the memory kernel from microscopic simulations \cite{Shin2010,carof2014two,jung2017iterative,meyer2020non}. Here, we employ the most straightforward reconstruction technique, directly based on the numerical 
inversion of the Volterra equation, 
	\begin{equation}\label{eq:Volterra2}
	\gj{ M	\frac{\ud}{\ud t}  {C}^\text{ss}_V(t) = - \int_{0}^t \ud s K(t-s) C^\text{ss}_V(s),}
	\end{equation}
	\gj{which is the stationary version of Eq.~(\ref{eq_volterra_general}).} 
Having reconstructed the memory kernel using time correlation functions determined in microscopic trajectories, we can directly use these trajectories to also access the fluctuating force via a trivial rewriting of the GLE,
	\begin{equation}
	\partial F(t) =  F(t) + \int_{0}^{t} \ud s K(t-s) V(s).
	\end{equation}
	Here, $ F(t) $ is the instantaneous force acting on the colloid,
    as calculated in the microscopic trajectory. This relations thus
    allows us to independently and \gj{unambiguously} verify the
    validity of the 2\FDTNS.  Importantly, it also enables us to access the probability distribution of $ \partial F(t) $.

\section{Computer simulations and modeling }
\label{sec:simulation_model}

In this work we simulate a colloid immersed in a DPD fluid. In DPD the fluid particles interact via dissipative and random pair forces, which are constructed such that the total momentum in the fluid
is conserved \cite{Hoogerbrugge_1992}. Both forces are connected via fluctuation-dissipation relations such that a canonical distribution is reached at equilibrium \cite{Espanol_1995}. The DPD equations of motion can be 
written as stochastic differential equations \cite{Espanol_1995}
\begin{align}
\ud \bm{r}_i =& \frac{\bm{p}_i}{m} \: \ud t \label{eq:dpdforce}\\
\ud \bm{p}_i =& \sum\limits_{j\neq i}^{} -\gamma \: \omega_\textrm{D}(\bm{r}_{ij})(\bm{e}_{ij}\cdot\bm{v}_{ij})\bm{e}_{ij} \ud t \\
&+  \sum\limits_{j\neq i}^{} \eta \: \omega_\textrm{R}(\bm{r}_{ij})\bm{e}_{ij} \ud W_{ij} \nonumber
\end{align}
with velocity difference $ \bm{v}_{ij} = \bm{v}_i - \bm{v}_j $,  distance $ \bm{r}_{ij} = \bm{r}_i - \bm{r}_j $, connection vector $ \bm{e}_{ij} = \bm{r}_{ij} / |\bm{r}_{ij} | $ and the 
fluctuation-dissipation relations $ \eta = \sqrt{2 k_\textrm{B} T \gamma} $ 
and $ \omega_\textrm{D}(r)=\omega_\textrm{R}(r)^2 $. The random forces are described by independent increments of a Wiener process $ \ud W_{ij} \ud W_{i^\prime j^\prime} = (\delta_{i i^\prime}\delta_{j j^\prime} + \delta_{i j^\prime}\delta_{j i^\prime}) \ud t $ \cite{Espanol_1995}. In the present work, we do not include any conservative forces in the DPD equations of motion. Since DPD is purely based on pairwise interactions, it can be regarded as a Galilean invariant
thermostat. Marsh \emph{et	al.} \cite{Marsh_1997} showed that DPD indeed reproduces the hydrodynamic equations
(Navier-Stokes equation) and calculated theoretical values for transport coefficients.

Here, we use the weight function
$\omega_\textrm{R}(r) = 1- \frac{r}{r_\textrm{cut}} $.
The simulation units are given by $k_B T = \epsilon$ (unit of energy), 
$r_\textrm{cut} = \sigma$ (unit of length length) and 
$ r_\textrm{cut} \sqrt{m/k_B T} = \tau $ (unit of time). 
We choose the DPD parameters, $ \gamma = 5\, \epsilon \tau \sigma^{-2} $, density $ \rho = 3\,\sigma^{-3} $ and the time step $ \Delta t = 0.005\tau $. The shear viscosity is $ \eta = 1.28\epsilon \tau \sigma^{-3}  $ \cite{Jung2016} and the solvent diffusion coefficient can be approximated to $ D_\textrm{s} =  0.75 \sigma^2 \tau^{-1} $ \cite{Marsh_1997}.

The colloid is modelled as a raspberry-like object, consisting of 80 independent particles placed on a spherical shell with radius $ R = 3\,\sigma. $ The total mass of the colloid is $M=80 m$.  The colloid is a rigid body, i.e.,
the relative distances of all particles forming the colloid are fixed.
These particles interact with the fluid particles via a
purely repulsive interaction, i.e., a truncated LJ potential with
cutoff $r_\textrm{c,LJ} = \sqrt[6]{2} \sigma$. We use a cuboid simulation box with periodic boundary conditions in all three dimensions and edge 
lengths $ L_x = 55.4689 \sigma $, $ L_y = L_z = 27.7345 \sigma $. To create the non-equilibrium steady-state we pull on the colloid with a constant and permanent force $ F^\textrm{ext} $ in positive $ x -$direction, and apply a very small negative bulk force to the fluid such that the total momentum in the system is conserved. All simulations are performed using
the simulation package {\em Lammps} \cite{Plimpton1995,Plimpton1995a}. 

To \gj{determine the response $\delta \langle V(t) \rangle$ and test
the 1\FDT in computer simulations}, we apply the perturbation $
\alpha(t) = M V_0 \delta(t) $ to a steady-state system. We then
perform two simulations in parallel; one with the perturbation (pert)
and one without (unpert). The response can then be calculated as $
\delta V(t) = V_\textrm{pert}(t) -V_\textrm{unpert}(t)  $. This
quantity is then averaged over many different systems, $ \delta
\langle V(t) \rangle $, with initial perturbations at $ t=0 $. With
this method, some statistical noise in the calculation of the response
function can be eliminated.

\section{Linear and non-linear response in active microrheology}
\label{Sec:Response}

In this Section we analyze the response of the colloid to the permanent external force, $ F^\textrm{ext}. $ After applying the force we simulate sufficiently long that the system reaches a steady state. All quantities that will be reported in the following are averages in these non-equilibrium steady states.

\subsection{Linear response}

\begin{figure}
	\includegraphics[scale=1]{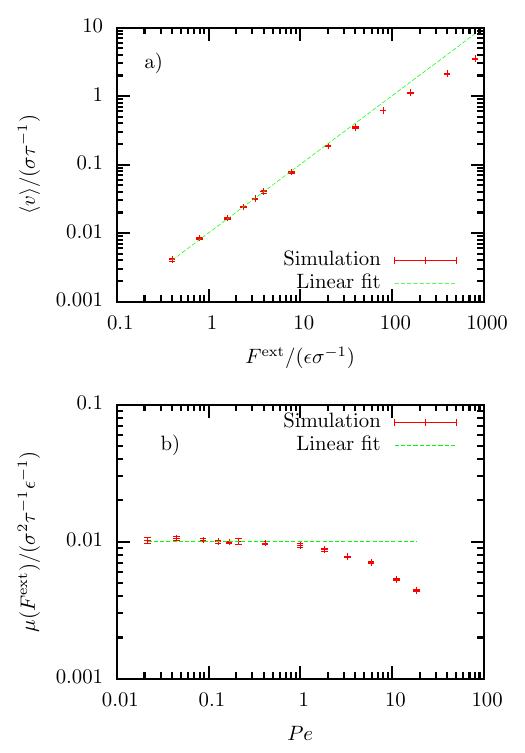}
	\caption{Response of colloid to an external force $ F^\textrm{ext}. $ in 
non-equilibrium steady-state. a): Average velocity $ \langle v \rangle $. 
b): Force-dependent mobility $ \mu(Pe) = \frac{\langle v \rangle}{F^\textrm{ext}}. $, plotted against Peclet number. The data was fitted in the linear regime, $ F^\textrm{ext} = [0,10] $, revealing an equilibrium mobility $ \mu = (0.0101 \pm 0.0001)\, \sigma^2 \epsilon^{-1}\tau^{-1}. $ 
 }
	\label{fig:linear_response}
\end{figure}

For small external forces we observe an extended linear response regime, in which the average steady-state velocity is given by, $ \langle v \rangle = \mu F^\textrm{ext}, $ with constant mobility $ \mu $ (see Fig.~\ref{fig:linear_response}a). Using the linear response regime, we can extract the mobility $ \mu = (0.0101 \pm 0.0001)\, \sigma^2 \epsilon^{-1}\tau^{-1}. $ An estimate of the mobility can also be determined using linear response theory, by integration of the VACF
\begin{equation}
\mu_{_\textrm{VACF}} = \frac{D}{k_B T} = \frac{1}{3 k_B T} \int_{0}^{\infty} \ud t C_V(t).
\end{equation}
and by integration over the memory kernel which appears in the generalized Langevin equation,
\begin{equation}
\mu_{_\textrm{K}} = \frac{1}{\gamma_c} = \left( \int_{0}^{\infty} \ud 
t K(t)  \right)^{-1}.
\end{equation}
The results for these dynamic correlation functions will be discussed later (see Section \ref{Sec:VACF_memory} and Fig.~\ref{fig:VACF_memory}). Extracting the mobility from these quantities results in $ \mu_{_\textrm{VACF}} = (0.0103 \pm 0.0001)\, \sigma^2 \epsilon^{-1}\tau^{-1}  $ and $ \mu_{_\textrm{K}} = (0.0113 \pm 0.0004)\, \sigma^2 \epsilon^{-1}\tau^{-1}   $ , in good agreement with the mobility determined above \footnote{The correlation functions were calculated up to a time $ t = 20 \tau $ and integrated numerically. To include the contributions of the long-time tails \cite{Hansen_Liquids2013} we fitted the long-time behaviour with the 
power law $ C_V(t) = A t^{-3/2} $ and integrated the contributions for $t > 20 \tau$ analytically.}. The discrepancy in the value $  \mu_{_\textrm{K}}  $ most probably arises from the memory reconstruction which becomes less accurate for longer times. Using Fourier transform techniques in the long-time regime as described in Refs.~\cite{Baity-Jesi2019,Hoefling2020} might improve these values.

Using the solvent diffusion coefficient, $ D_s $, we can also define the Peclet number, $ Pe = \langle v \rangle / v_\textrm{diff} $, with $ v_\textrm{diff} \approx \frac{D_s}{R + \sigma} \approx 0.19 \sigma \tau^{-1} 
$. This dimensionless quantity thus quantifies the ratio of the advective 
transport due to the external force to the diffusive transport.  The linear response regime extends to Peclet numbers of roughly $ Pe < 1 $, as can be observed in Fig.~\ref{fig:linear_response}b. For larger driving forces, the mobility clearly depends on the strength of the external force. 

\subsection{Beyond linear response: Thickening}

Beyond the linear response regime, different non-linear behaviours
have been observed, including thinning in Brownian suspensions
\cite{Brady2005,Puertas_2014} and glass-forming Yukawa fluids
\cite{Horbach2012,Harrer_2012}, as well as thickening in granular
systems \cite{Buchholtz1997InteractionOA,Wang2014,Puertas_2014}. Both
thinning and thickening behaviour has been observed in a model
colloidal system with solvent particles described by a Langevin
equation \cite{Sperl2016_AM}. The authors explain this with the
transition between a diffusive and a damping regime a low Peclet
numbers and from the damping to a collision regime at high Peclet
numbers.  In our simulations using a dense DPD fluid, we do not
observe a thinning regime, but directly thickening at $ Pe > 1, $ (see
Fig. \ref{fig:linear_response}b). 

\section{Structure, fluctuations and dissipation in the colloid frame}
\label{Sec:FDR}

Having analysed the velocity which the colloid attains in the non-equilibrium steady state we will now study its properties in the colloid frame. This includes density and velocity profiles of the surrounding fluid, as well as validations of the two fluctuation-dissipation relations as introduced in Section \ref{Sec:Methods}.

\subsection{Radial distribution function and velocity profiles}

\begin{figure}
	\centering\includegraphics[scale=0.4]{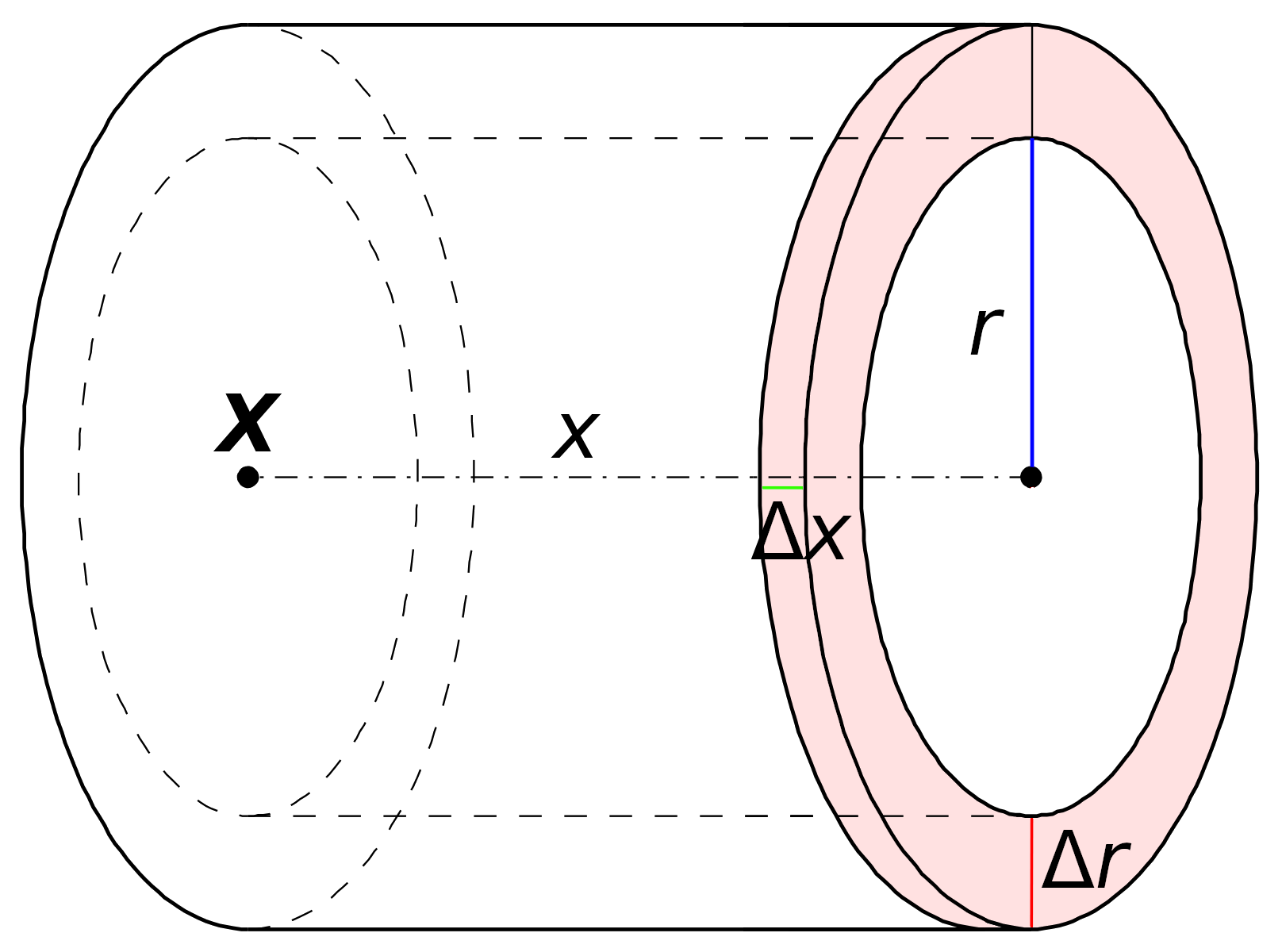}
	\caption{Cylindrical geometry in which the profiles in Fig.~\ref{fig:static} are evaluated. The colloid center is $ \bm{X} $. In this work we have used $ \Delta r = 0.2\,\sigma  $ and $ \Delta x = 0.4\,\sigma. $ }
	\label{fig:geometry}
\end{figure}

\begin{figure*}
	
	\includegraphics[scale=1]{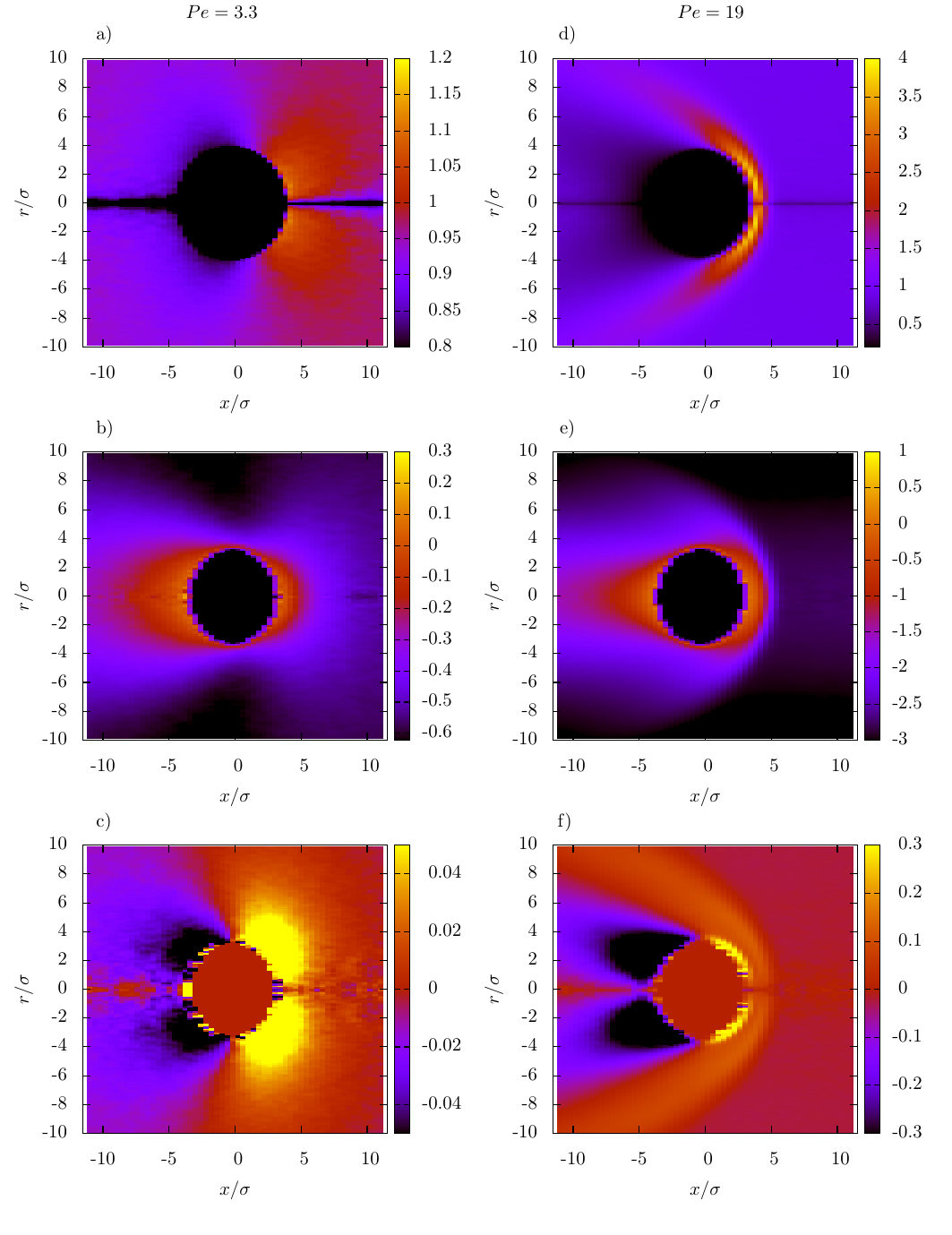}
	
	\vspace*{-0.8cm}
	
	\caption{ Radial distribution function and velocity profiles of DPD particles around the colloid in non-equilibrium steady-states with different Peclet numbers $ Pe $. The quantities are calculated in a cylindrical geometry where the axis is set by the direction of the external force (pulling in positive $ x $-direction). Averages are taken in wheel-shaped slices with radius $ r $ and a mid-point distance $ x $ from the colloid center (see Fig.~\ref{fig:geometry}). For the sake of visualization, the profiles are mirrored at $ r=0 $, i.e. $ f(-r) = f(r) $. Shown are radial distribution function $ g(x,r) $ a),d), average velocity parallel b),e) and perpendicular c),f) to the direction of the external field. The stripe 
at $ |r| < 0.2\sigma $ results from insufficient sampling in a very small 
volume. The colour code shows the magnitude of the scalar fields (different scales for each plot).  }
	\label{fig:static}
\end{figure*}

To quantify the density profile around the colloid we calculate the radial distribution function,
\begin{equation}
g(x,r) = \left\langle \frac{1}{2\pi r \rho \Delta x} \sum_{j:x_j\in[x,x+\Delta x]} \delta(|\bm{R} - \bm{r}_j| -r), \right\rangle,
\end{equation}
using the cylindrical geometry sketched in Fig.~\ref{fig:geometry}. Depending on the Peclet number, the radial distribution functions behave qualitatively different (see Fig.~\ref{fig:static}a,d). While for smaller Peclet number the structural deformation still reminds of a diffusive dipole \cite{Brady2005b} (albeit $ Pe $ is already relatively large in \ref{fig:static}a) the structure is completely different for large $ Pe $ in which 
a significant bow wave emerges and a wake with much fewer particles trails the colloid. This quantitative difference between $ Pe \lessapprox 3.3 $ and $ Pe \gtrapprox 18.6 $ is also perfectly visible in the velocity profiles. While for small $ Pe $ the profiles still display a certain symmetry between the front and the back of the colloid (see Fig.~\ref{fig:static}b,c), for higher $ Pe $ this symmetry is broken. In particular, the bow wave in front of the colloid is nicely visible in Fig.~\ref{fig:static}e and the wake behind the colloid in Fig.~\ref{fig:static}f.

From this analysis we can thus conclude that although the linear response regime only extends up to $ Pe < 1 $, the properties of the surrounding fluid significantly changes only for larger $ Pe > 3.3 $. We therefore expect that this similarly holds for the dynamic properties of the colloid in the colloid frame.

\begin{figure*}
	\includegraphics[scale=1]{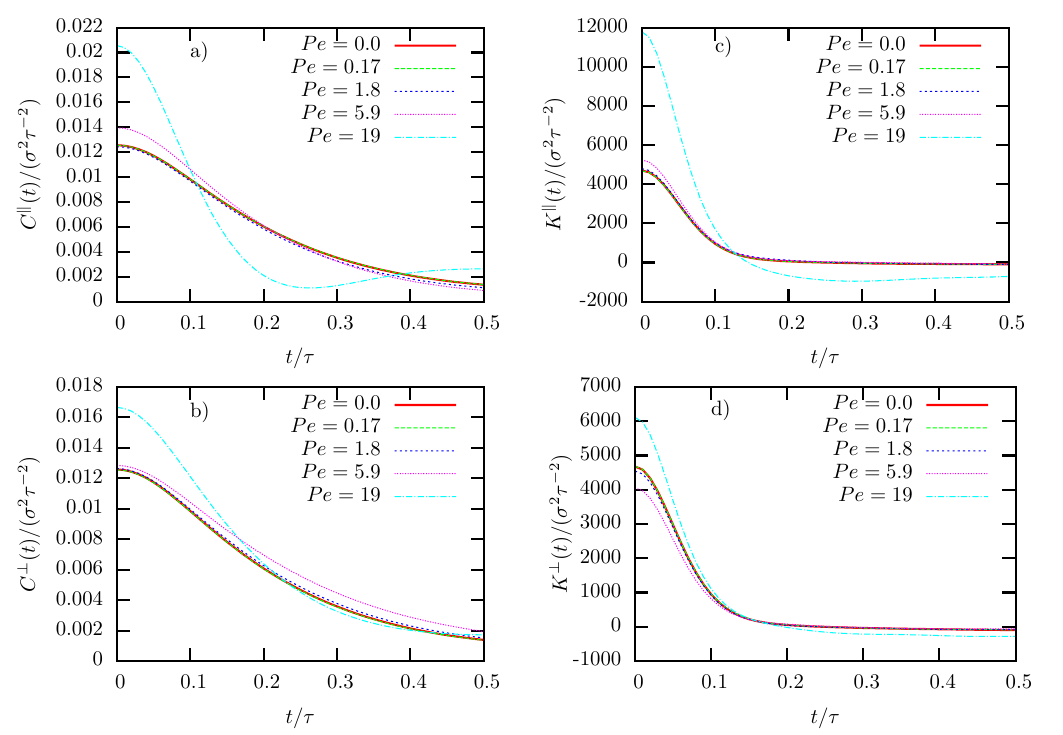}
	\caption{Velocity auto-correlation function $ C(t) $ and memory kernel $ 
K(t) $ of colloids in non-equilibrium steady-states with different Peclet 
numbers $ Pe $. Velocity fluctuations and memory kernel are calculated in 
the colloid frame parallel a),c) and perpendicular b),d) to the direction 
of the external force.  }
	\label{fig:VACF_memory}
\end{figure*}

\subsection{Dynamic correlations and memory kernels}
\label{Sec:VACF_memory}

The velocity auto-correlation function (VACF) of the colloid in the colloid frame is shown in Fig.~\ref{fig:VACF_memory}. Without external driving, the VACF is governed by an initial exponential decay, followed by the usual hydrodynamic ling-time tail, which can be described by the power law 
$ C_V(t) \sim t^{-3/2} $ \cite{Hansen_Liquids2013,Zhou_2016} (not shown here due to large statistical fluctuations). As expected from linear response theory, the correlation functions for $ Pe < 1 $ are independent of the external force and isotropic. When increasing the driving above $ Pe > 
1 $ the first small deviations are observable for larger times, which are, however, barely visible, in agreement with our previous observations. For very large Peclet number the VACF then qualitatively changes. First, we observe an increase in ``local temperature'' as defined
kinetically via $T_\text{neq} \sim C_V(t=0)$. Second, the changes of the local fluid structure induce an oscillatory behavior in the VACF parallel to the external driving, $ C^\parallel(t) $, as can be seen in Fig.~\ref{fig:VACF_memory}a. If the colloid moves in the negative x-direction, it leaves the bow wave which counteracts the external driving, which automatically means that the external force will push the colloid back into position. If the colloid moves in the positive $ x $-direction, the density 
of fluid particle significantly increases which similarly leads to a restoring force. Both cases thus effectively induce a ``trapped'' motion, which explains the oscillations in the VACF. In perpendicular direction, this effect is much smaller, most importantly, the local temperature does not increase as much as in parallel direction, i.e. $ T_\textrm{neq}^\perp < T_\textrm{neq}^\parallel  $  (see Fig.~\ref{fig:VACF_memory}b). Moreover, in the intermediate driving regime at $ Pe = 5.9 $ one in fact observes a slower decay of the VACF, which is most probably due to the small decrease in density in the direction perpendicular to the colloid. Only when increasing the driving even further, one observes a similar behavior as discussed in parallel direction, consistent with the change in structure shown in Fig.~\ref{fig:geometry}a,d.

 The same observations hold for the memory kernel $ K(t) $. In equilibrium, the memory kernel also has an initial exponential decay, however for larger times it becomes negative and approaches zero with the same power law as the VACF but different sign, $ K(t) \sim - \frac{\gamma_c^2}{M} \frac{C_V(t)}{C_V(0)} $ \cite{Corngold1972,Hoefling2020}. The oscillatory dependence on $ t $ discussed in the VACF for large $ Pe $ is reflected in the memory kernel by a strong initial damping, followed by a very pronounced minimum with negative friction (see Fig.~\ref{fig:VACF_memory}c,d).

\subsection{Violation of 1\FDT for strong external driving}
\label{sec:1FDR_vio}

\begin{figure}
	\includegraphics[scale=1]{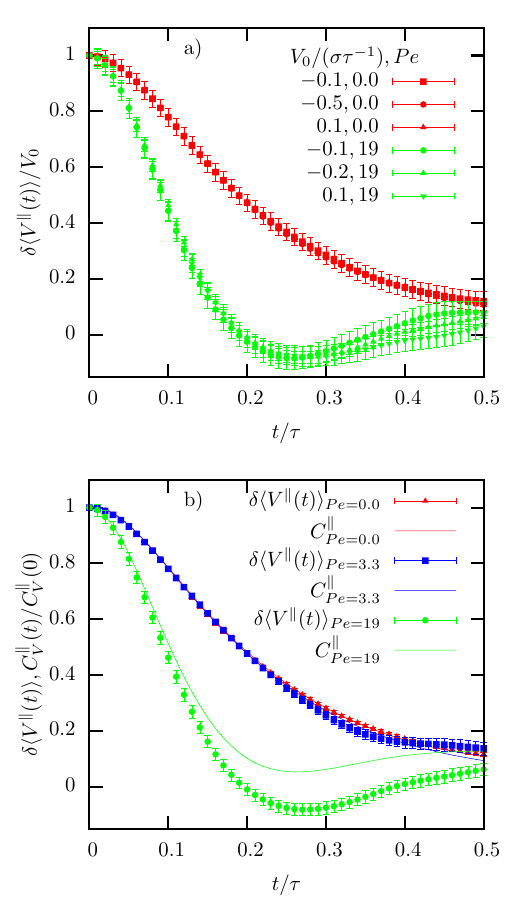}
	\caption{Non-equilibrium response to a force impulse at time $t=0$ in non-equilibrium steady-states with different Peclet numbers $ Pe $. The perturbation $\alpha(t) = M V_0 \delta(t)$ is applied parallel to the permanent external force.  a) Response $ \delta \langle V^\parallel(t) \rangle $, for different values of  the initial velocity difference $ V_0 $. b) Response compared to the normalized velocity auto-correlation function 
$ C_V^\parallel(t)/C_V^\parallel(0) $, evaluated in steady-state. }
	\label{fig:1FDR}
\end{figure}

In the colloid frame, we can also calculate the non-equilibrium response $ \delta \langle V^\parallel(t) \rangle$ of the colloid to a perturbation 
induced by a force impulse at time $ t=0$ in the positive $x$-direction, as defined in Section \ref{Sec:Linear_response}. We emphasize that this 
force impulse is applied {\em in addition} to the permanent external force $F_\text{ext}$. Hence we consider a time-dependent perturbation in a stationary non-equilibrium state in which the colloid moves with a constant 
velocity, driven by the permanent external force.  In equilibrium systems, according to the linear response theory, this response depends linearly 
on the amplitude of the perturbation. As can be seen in Fig.~\ref{fig:1FDR}a, the normalized response is independent of the strength and direction 
of the force impulse (i.e. parallel or anti-parallel to the external driving of the colloid), which shows that this expectation is still fulfilled 
at nonequilibrium.

Comparing this normalized response to the VACF, we can immediately
investigate the validity of the 1\FDTNS. While in equilibrium the 1\FDT is
indeed fulfilled, for strong external forces we can clearly observe a
strong violation of the first fluctuation-dissipation relation. To
rationalize this important observation, we recall the results of the
previous Section. There, we have discussed that the instantaneous
fluctuations of the velocity, $ C_V(0), $ in the directions parallel
and perpendicular to the external force are significantly different. A
closer look at the bow wave in Fig.~\ref{fig:static}d,e,f also shows
that the structure in the surrounding fluid can induce a coupling
between these two different directions. This means that the effective
restoring forces, $ F^\parallel(x,r) $, which induce the oscillatory
behavior of $ C_V^\parallel(t) $ at strong driving, do not only depend
on $ x $, but also on $ r $ (and similarly in perpendicular
direction). We therefore have precisely the situation described in the
introduction, with two coupled degrees of freedom which have different
temperature\cite{Villamaina2009_FDTTHERMO,Villamaina2012_FDTTHERMO}.
Here, the situation is clearly more complicated than in this toy model
and it is not possible to disentangle the different contributions to
the response function, but the model gives a reasonable explanation
for the (apparent) violation of the 1\FDT in active microrheology.

\subsection{Thermal fluctuations and 2\FDT }

Having investigated the first fluctuation-dissipation relation, we now use the methodology described above (see Section \ref{ref:MZ}) to determine 
the thermal fluctuations in active microrheology and thus the second fluctuation-dissipation relation.

\begin{figure}
	\includegraphics[scale=1]{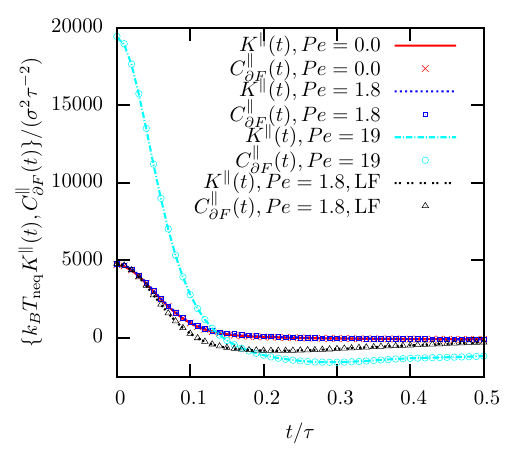}
	\caption{Memory kernels, $k_B \gj{T_\text{neq}} K^\parallel(t)$, and auto-correlation function of the stochastic force, $C^\parallel_{\partial F}(t)$, in non-equilibrium steady-states with different Peclet numbers $ Pe 
$. The last two curves (black) correspond to the application of the memory kernel formalism in the laboratory frame (LF).   }
	\label{fig:2FDR}
\end{figure}

Comparing the time-correlation function of the thermal forces, $
C_{\partial F}(t) = \left\langle   \partial F(0)  \partial F(t)
\right\rangle_\text{ss} $ with the memory kernel $ K(t) $ discussed at
the beginning of this Section, we clearly see that the 2FDT is
fulfilled for all different driving forces (see Fig.~\ref{fig:2FDR}).
Different from the 1\FDTNS, which only holds strictly in equilibrium
conditions (at least in its naive version, as discussed in
Ref.~\cite{Villamaina2009_FDTTHERMO}), the 2FDT indeed remains valid
in a non-equilibrium steady state. \gj{This numerically confirms the
theoretical calculations presented in Section \ref{Sec:Methods} and
derived in Appendix \ref{app:2fdr}.} Interestingly, one can also
calculate the correlation functions in the laboratory frame and
extract the memory kernel and the thermal fluctuations in the same way
as described before. The resulting memory kernel will clearly be
different, but the validity of the 2FDT is not affected (see
Fig.~\ref{fig:2FDR}, black curve). This result also highlights the
importance of describing the system in the \gj{colloid} frame. While
both descriptions are mathematically sound, the description in the
colloid frame highlights the universality of the memory kernel inside
the linear response regime (and even beyond). 

\begin{figure}
	\includegraphics[scale=1]{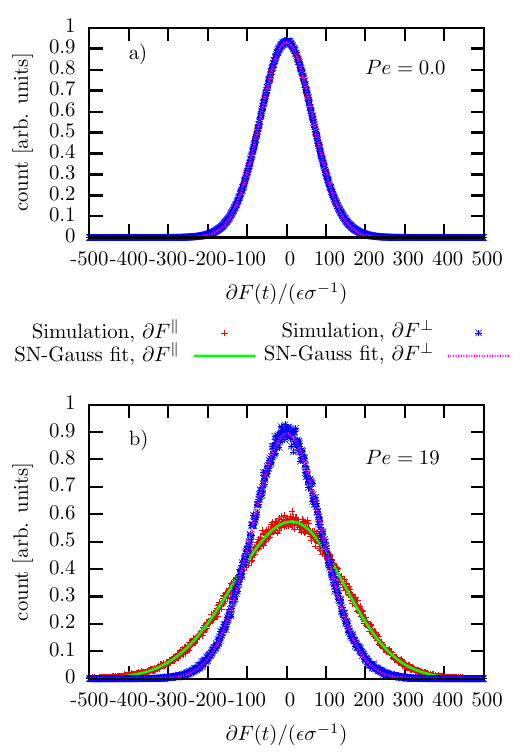}
	\caption{Distribution of the stochastic force, $ \partial F(t) $ in non-equilibrium steady-states with different Peclet numbers $ Pe $. The data is fitted with a split normal distribution (see Eq.~(\ref{SN_gauss})) (SN-Gauss). In the top panel, all curves perfectly overlap.  }
	\label{fig:SN_gauss}
\end{figure}

We also investigate the distribution of the thermal fluctuations. In
equilibrium, the distribution is an almost perfect Gaussian function,
as one might expect from central limit theorem (see
Fig.~\ref{fig:SN_gauss}a), since the total force consists of hundreds
of collisions which are basically independent (apart from hydrodynamic
interactions).  Interestingly this no longer holds outside the linear
response regime. In Fig.~\ref{fig:SN_gauss}b one can clearly observe a
slight asymmetry in the distribution of forces parallel to the
external driving, which occurs due to a long tail of ``negative
forces'' (i.e. anti-parallel to the external force). This tail emerges
since the colloid is constantly pulled through an otherwise stationary
fluid. Inside the linear response regime, the diffusive dipole as
discussed in Ref.~\cite{Brady2005b} ensures that the particles close
to the colloid indeed have the same relative velocity (and thus the
same statistics of collisions). This is no longer the case for $ Pe >1
$, hence some fluid particles, illustratively, crash into the colloid,
and thus induce large negative forces. Since the total average force
is zero, these strong negative forces have to be compensated by a
slightly enhanced probability of observing a positive thermal force.
The distribution outside the linear response regime can, in fact, be
described by a split normal distribution (SN-Gauss),
\begin{equation}\label{SN_gauss}
f(x) = A \left\{ \begin{array}{ll}
e^{-\frac{(x-f_\textrm{max})^2}{2 \sigma_L^2}} & x < f_\textrm{max}\\
e^{-\frac{(x-f_\textrm{max})^2}{2 \sigma_R^2}} & x \geq  f_\textrm{max} 
\end{array} \right.
\end{equation}
with mean $ \bar{f} = 0 $ and maximum $ f_\textrm{max} = \sqrt{\frac{2}{\pi}} (\sigma_L - \sigma_R) > 0. $

\begin{figure}
	\includegraphics[scale=1]{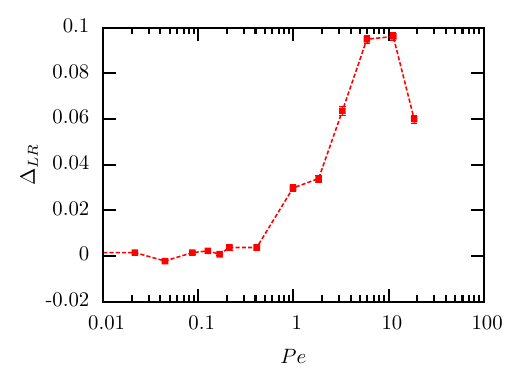}
	\caption{Asymmetry factor $ \Delta_{LR} $ of the split normal distribution as fitted in Fig.~\ref{fig:SN_gauss}. The error bars are smaller than or basically have the same size as the points.  }
	\label{fig:asymmetry}
\end{figure} 

Using the split normal distribution, we can define an asymmetry factor
$ \Delta_{LR} = \frac{\sigma_L - \sigma_R}{\sigma_L + \sigma_R} $. It
shows an unexpected non-monotonic dependence on the $ Pe $ number (see
Fig.~\ref{fig:asymmetry}). For $ Pe < 1 $ it is clearly zero,
consistent with the above discussion of the linear response regime.
Intriguingly, we can observe a very sharp transition away from
$\Delta_{LR}=0$, allowing us to determine the end of the linear
response regime with much more precision than possible from a simple
inspection of the average steady-state velocity. Furthermore, the
asymmetry reaches a maximum at around $ Pe = 5 $ and then decays
rapidly again. We explain this behavior with the formation of a thick
and dense particle ``shield'' as illustrated in Fig.~\ref{fig:static}.
This bow wave is dense enough to efficiently accelerate the particles
in front of the colloid, shielding it from stronger impacts, as
described above.

\section{Discussion}
\label{Sec:Discussion}

In this paper we have investigated the validity and potential
violations of fluctuation-dissipation relations in a driven system far
from equilibrium.  We found that the 1\FDT is only valid under very
restrictive conditions. On the other hand, we provided a mathematical
argument and numerical evidence that the 2\FDT is exactly fulfilled
for all values of driving forces, even far beyond the non-linear
response regime. 

As mentioned in the introduction, violations of the 2\FDT have
repeatedly been reported in the literature \cite{Falasco2014,
Maes2014, Steffenoni2016, Srivastava2018, Zaccone2018_FDTNESS}. The
reason is that, when investigating the forces on a selected probe
particle due to an orthogonal bath, it is not {\em a priori} clear how
to distribute them between the memory and the noise term in the GLE,
without additional requirements.  Mitterwallner \emph{et al.}
\cite{Netz2020_FDTNESS} have recently pointed out that an infinite
number of GLEs are compatible with a given VACF $C^\text{ss}_V(t)$.
However, if one imposes an orthogonality condition \fs{on the noise}
or, equivalently, the validity of the Volterra equation, this singles
out one GLE, in which the 2\FDT is fulfilled.

This remains correct in non-stationary situations, \fs{as discussed in
Section \ref{ref:MZ}} and should also be valid in the presence of
(external) drift terms, i.e., for GLEs of the form
 \begin{equation}
   \label{eq:gle_drift}
        M \frac{\ud}{\ud t} {V}_i(t) = f_i(t) - \int_{T}^t \!\! \ud s \: 
         K_{ij}(t,s) \: V_j(s) + \partial F_i(t).
\end{equation}
For a given ensemble of trajectories, it can be constructed {\em via} the 
following steps:
    \begin{enumerate}
        \item Determine $V^0_i(t) = \langle V \rangle^\text{neq}_i(t)$.
        \item Rewrite $V_i(t) = V^0_i(t) + u_i(t)$,
        determine $C^u_{ij}(t,t_0) = \langle u_i(t) u_j(t_0) \rangle_\text{neq}$ and
        then Eq. (\ref{eq:gle_drift}) can be separated as follows
        \begin{align*}
         M\dot{V}^0_i =& f_i(t) - \int_{T}^t \!\!\! \ud s \:
             K_{ij}(t,s) \: V^0_j(s) \\
         M\dot{u}_i(t) =&  - \int_{T}^t \!\!\! \ud s \:
             K_{ij}(t,s) \: u_j(s)  + \partial F_i(t).
        \end{align*}
        \item Determine $K_{ij}(t,t_0)$ by Volterra inversion of $C^u_{ij}(t,t_0)$.  
        One obtains a GLE for $u_i(t)$ that satisfies the 2\FDTNS.
        \item Determine the effective drift force {\em via} 
        $f_i(t) = \dot{V}^0_i(t) + \int_{T}^t \ud s \:K_{ij}(t,s) \: V^0_j(s)$.
    \end{enumerate}
The resulting GLE would satisfy the 2\FDTNS.

Cui \emph{et al.} have discussed a particularly intriguing case of a
particle coupled to a bath of charged oscillators and subject to an
oscillatory electric field \cite{Zaccone2018_FDTNESS}. They  derived a
GLE by integrating out the bath particles following a procedure
outlined by Zwanzig \cite{Zwanzig2001}. The resulting GLE does not
satisfy the 2FDT, moreover, the noise has a deterministic component
that reflects the oscillatory motion of the charged bath particles.
Based on our construction above we argue that also in this system, an
equivalent GLE can be constructed that does satisfy the 2\FDTNS.

To summarize, it is possible to formulate GLEs that do not satisfy the
2\FDTNS. It some cases, working with them may be more convenient -- they
may have a simpler  structure or the simulation may be easier.
However, \gj{different from the 1\FDT} one cannot use such equations
to postulate a {\em fundamental} violation of the 2\FDTNS, as it is
always possible to construct equivalent GLEs that do satisfy the
2\FDTNS. If the GLE is constructed based on the principle that the noise
should be perpendicular to the selected variable at some time $t=T$,
then this automatically results in the 2\FDT relation. From a
modelling perspective this latter choice strikes us as desirable since
it enables a systematic and unique separation into deterministic
drift, deterministic memory and friction forces as well as stochastic
noise, as illustrated above. 

\section{Conclusion}
\label{Sec:Conclusion}

In this work we have investigated the dynamical properties of colloids in
a system far from equilibrium, in which a colloid is pulled with a
constant force through a fluid. First, we have identified the linear
response regime and characterized the shear thickening behaviour of
the suspension when driving the colloid beyond linear response.
Second, we have investigated dynamic properties in the Galilean
reference frame which moves with the average velocity of the colloid.
We were thus able to characterize in detail the impact of the
non-equilibrium conditions on the dynamic correlation functions, the
memory kernels and the \FDTsNS.

With our analysis, we have observed a violation of the 1\FDT, i.e. the
relationship between non-equilibrium response and the stationary
correlation function. The violation can be explained by the emergence
of two different ``temperatures'' in the direction parallel and
perpendicular to the external driving. Furthermore, we have validated
the 2\FDTNS, i.e. the connection between the dissipative and stochastic interaction in the system, even in conditions far away from equilibrium.  We have further studied the properties of the stochastic forces and found 
an emerging asymmetry in its distribution function, which can be described by a screw normal distribution. This asymmetry appears to be a strong indicator to determine the linear response regime, since it depends very sensitively on perturbations of the usual diffusive dipole \cite{Brady2005b}.

The purpose of this work is to engage a discussion on
fluctuation-dissipation relations, particular the 2\FDTNS, in out-of-equilibrium conditions. As we have argued in the previous Section and as has been argued by Mitterwallner \emph{et al.} \cite{Netz2020_FDTNESS}, the distinction between systematic and stochastic
interactions with bath particles is {\em a priori} somewhat arbitrary. 

We therefore suggest to impose, as additional fundamental criterion, an 
orthogonality condition as it directly follows from the Mori-Zwanzig
formalism \cite{Zwanzig2001,Meyer2017}. This \emph{uniquely} defines
the relationship between the dissipative and the random forces in the
system, which is then given by the 2\FDTNS, and it is applicable to
systems far away from equilibrium and also non-stationary dynamics.
Such a separation is crucial for consistent modeling and should enable
to use dynamic coarse-graining techniques developed in recent years
\cite{Li2015,Jung2018,wang2020data} for non-equilibrium systems. From
a practical point of view, it could sometimes be convenient to use
equivalent versions of the GLE that violate the 2\FDTNS. However, in our opinion, this should then be seen as a mathematical trick rather than a fundamental property of the underlying dynamical system.

\section*{Acknowledgements}
The authors thank Thomas Speck, J\"urgen Horbach, Martin Hanke, Jeanine Shea and Thomas Franosch for helpful discussions, as well as Andre Gladbach for his contributions in an early stage of this project. This work has been supported by the DFG within the Collaborative Research Center TRR 146 via Grant 233630050, project A3. GJ also gratefully acknowledges funding by the Austrian Science Fund (FWF): I 2887.

\appendix

\section{Volterra Equation, Orthogonality, and 2\FDT in Generalized Langevin equations}
\label{app:2fdr}
We consider a general non-stationary GLE of the form
\begin{equation}
\label{eq:gle_general2}
\gj{ \frac{\ud }{\ud t} A_i(t) =  - \int_{T}^{{t}} \ud s K_{i k}(t,s) A_k(s) + \partial F_i(t),}
\end{equation}
where the memory kernel $K_{i k}(t,s)$ is not required to be invariant with respect to time translation. Likewise, the correlation function $C_{ij}(t,t_0) = \langle A_i(t) A_j(t_0) \rangle_\text{neq}$ is not assumed to necessarily depend on $(t-t_0)$ only. 
The time $T$ is some arbitrary ``initial'' time, $T<t$, which can also
be chosen $T \to -\infty$.

In the following, we will discuss the connection between the three following relations:
\begin{itemize}
    \item[(I)] The deterministic Volterra equation
    \begin{equation}
\gj{          \frac{\ud}{\ud t} C_{ij}(t,t_0) = 
         - \int_{t_0}^t \ud s \: K_{i k}(t,s) \: C_{k j}(s,t_0)  }
    \end{equation}
    \item[(II)] An expression for the correlation between the random force and velocity
    \begin{equation}
\gj{\langle \partial F_i(t) \: A_j(t_0) \rangle_\text{neq}
= \int_T^{t_0} \ud s \: K_{i k}(t,s) \: C_{kj}(s,t_0) }
    \end{equation}
    for all time pairs $(t,t_0$)
    \item[(III)] The second fluctuation-dissipation relation
    \begin{equation}
\gj{        \langle \partial F_i(t) \: \partial F_j(t_0) \rangle_\text{neq}
           =  K_{ik}(t,t_0) \: C_{kj}(t_0,t_0)}
    \end{equation}
\end{itemize}

The relation (I) and (III) have been shown to hold for non-stationary Hamiltonian systems using the Mori-Zwanzig formalism \cite{Meyer2017}\footnote{The relation (III) corresponds to Eq.\ (39) in Ref.\ \cite{Meyer2017}, 
written in the form of a Taylor expansion.}.  

\gj{We can easily see that (I) and (II) are equivalent. We simply
multiply  Eq.\ (\ref{eq:gle_general2}) with $A_j(t_0)$ and take the
ensemble average. We will now show that (I) implies (III). To this
end, we first express $\partial F_j(t_0)$ using Eq.\
(\ref{eq:gle_general2}) multiply with $\partial F_i(t)$, take the
ensemble average and thus write the force-force correlation as
\begin{align} 
\label{eq:d1} 
\langle \partial F_i(t) &\: \partial F_j(t_0) \rangle_\text{neq} 
  = \langle \partial F_i(t) \dot{A}_j(t_0) \rangle_\text{neq} \\ 
  \nonumber & + \int_T^{t_0} \ud s \: K_{jk}(t_0, s) \: 
  \langle \partial F_i(t)  A_k(s)  \rangle_\text{neq}.
\end{align} 
For the first term on the right hand side we find
\begin{align} 
\label{eq:d2} 
\langle \partial F_i(t)  \: \dot{A}_j(t_0)  \rangle_\text{neq} 
= K_{i k}(t,t_0)C_{kj}(t_0,t_0) + \nonumber \\ 
 \int_T^{t_0} \ud s \: K_{i k}(t,s) \: 
  \frac{\ud}{\ud t_0} {C}_{kj}(s,t_0)  \rangle 
  \end{align} 
which is obtained by taking the derivative of (II) with respect 
to $t_0.$ The second term can be rewritten as 
\begin{align} 
\label{eq:d3} 
\lefteqn{\int_T^{t_0}\!\!\!  \ud s \: K_{jk}(t_0, s) \: \langle 
 \partial F_i(t)  A_k(s) \rangle_\text{neq}} \quad && \\ 
 \nonumber & =  \int_T^{t_0} \!\!\! \ud s \: K_{jk}(t_0,s) \: 
 \int_T^s \!\!\! \ud s' \: K_{il}(t,s') \: C_{lk}(s',s) \\ 
 \nonumber & = \int_T^{t_0} \!\!\! \ud s'\: K_{il}(t,s') \: 
   \int_{s'}^{t_0} \!\!\! \ud s \: K_{jk}(t_0,s) \: C_{kl}(s,s') \\ 
    \nonumber & = -  \int_T^{t_0} \!\!\! \ud s'\: K_{il}(t,s') \: 
    \frac{\ud}{\ud t_0} C_{jl}(t_0, s').\\ 
    \nonumber &= - \int_T^{t_0} \!\!\! \ud s'\: K_{ik}(t,s') \: 
    \frac{\ud}{\ud t_0} C_{kj}(s',t_0) 
\end{align} 
Here, we have used (II) in the first step, the symmetry property
$C_{ij}(s',s) = C_{ji}(s,s')$ in the second step and (I) in the last
step.  Combining Eqs.\ (\ref{eq:d1}, \ref{eq:d2}, \ref{eq:d3}), we
finally obtain the fluctuation dissipation relation (III). Hence the
second fluctuation dissipation relation is a necessary consequence of
the deterministic Volterra equation. We emphasize that this is a
general relation, which does not rely on the Mori-Zwanzig formalism.}

Based on these general results, we can now specifically
discuss stationary GLEs with \gj{$K_{ij}(t,s) = K_{ij}(t-s)$} and their
stationary solutions with \gj{$C_{ij}(t,t_0) = C_{ij}(t-t_0)$}.
We consider the two most popular such GLEs, the Mori-Zwanzig
GLE with $T=0$, and the so-called ``stationary GLE''
with $T \to -\infty$ \cite{Shin2010}. 
In these cases, the conditions (I-III) read
\gj{
\begin{equation}
\hspace*{-1.6cm}(I) \quad \frac{\ud}{\ud t} C_{ij}(t) = 
- \int_{0}^t \ud s \: K_{ik}(t-s) \: C_{kj}(s)
\label{eq:volterra_app}
\end{equation}
\begin{itemize}
 \item[(\emph{II})]  \mbox{(Mori-Zwanzig GLE):} 
 $
 \langle   A_i(0) \partial F_j(t)\rangle_\text{ss} = 0
$
\item[(\emph{II'})]  \mbox{(Stationary GLE) \cite{Shin2010}:}
\begin{equation}\label{key}
 \langle   A_i(0) \partial F_j(t) \rangle_\text{ss} = 
  \int_{-\infty}^0 \ud s \: K_{ik}(t-s) C_{kj}(s) 
\end{equation}
\item[(\emph{III})] $  \langle  \partial F_i(0) \partial F_j(t) \rangle_0
           =  K_{ik}(t) \: C_{kj}(0)$
\end{itemize}
}


\fs{ We close with a note on the invertibility of Eq.
(\ref{eq:volterra_app}). Provided $C_{ij}(t)$ has a locally integrable
second derivative, Eq. (\ref{eq:volterra_app}) can be inverted in a
straightforward manner by Fourier methods, yielding a unique memory
kernel $K_{ij}(t)$.  Whereas this integrability condition is usually
met at $t > 0$, it can broken at $t=0$ if $\dot{C}_{ii}(t) \neq 0$ at
$t \to 0^+$ (since $\dot{C}_{ii}(t) = - \dot{C}_{ii}(-t)$).  However,
such cases can be handled as well. The memory kernel then acquires a
$\delta$-shaped instantaneous friction contribution.  }

\clearpage

\bibliography{library}

\end{document}